\PassOptionsToPackage{unicode}{hyperref}
\PassOptionsToPackage{hyphens}{url}
\PassOptionsToPackage{dvipsnames,svgnames,x11names}{xcolor}
\documentclass[
  12pt]{article}

\usepackage{amsmath,amssymb}
\usepackage{iftex}
\ifPDFTeX
  \usepackage[T1]{fontenc}
  \usepackage[utf8]{inputenc}
  \usepackage{textcomp} % provide euro and other symbols
\else % if luatex or xetex
  \usepackage{unicode-math}
  \defaultfontfeatures{Scale=MatchLowercase}
  \defaultfontfeatures[\rmfamily]{Ligatures=TeX,Scale=1}
\fi
\usepackage{lmodern}
\ifPDFTeX\else  
\fi
\IfFileExists{upquote.sty}{\usepackage{upquote}}{}
\IfFileExists{microtype.sty}{% use microtype if available
  \usepackage[]{microtype}
  \UseMicrotypeSet[protrusion]{basicmath} % disable protrusion for tt fonts
}{}
\makeatletter
\@ifundefined{KOMAClassName}{% if non-KOMA class
  \IfFileExists{parskip.sty}{%
    \usepackage{parskip}
  }{% else
    \setlength{\parindent}{0pt}
    \setlength{\parskip}{6pt plus 2pt minus 1pt}}
}{% if KOMA class
  \KOMAoptions{parskip=half}}
\makeatother
\usepackage{xcolor}
\makeatletter
\ifx\paragraph\undefined\else
  \let\oldparagraph\paragraph
  \renewcommand{\paragraph}{
    \@ifstar
      \xxxParagraphStar
      \xxxParagraphNoStar
  }
  \newcommand{\xxxParagraphStar}[1]{\oldparagraph*{#1}\mbox{}}
  \newcommand{\xxxParagraphNoStar}[1]{\oldparagraph{#1}\mbox{}}
\fi
\ifx\subparagraph\undefined\else
  \let\oldsubparagraph\subparagraph
  \renewcommand{\subparagraph}{
    \@ifstar
      \xxxSubParagraphStar
      \xxxSubParagraphNoStar
  }
  \newcommand{\xxxSubParagraphStar}[1]{\oldsubparagraph*{#1}\mbox{}}
  \newcommand{\xxxSubParagraphNoStar}[1]{\oldsubparagraph{#1}\mbox{}}
\fi
\makeatother

\providecommand{\tightlist}{%
  \setlength{\itemsep}{0pt}\setlength{\parskip}{0pt}}\usepackage{longtable,booktabs,array}
\usepackage{calc} % for calculating minipage widths
\usepackage{etoolbox}
\makeatletter
\patchcmd\longtable{\par}{\if@noskipsec\mbox{}\fi\par}{}{}
\makeatother
\IfFileExists{footnotehyper.sty}{\usepackage{footnotehyper}}{\usepackage{footnote}}
\makesavenoteenv{longtable}
\usepackage{graphicx}
\makeatletter
\def\maxwidth{\ifdim\Gin@nat@width>\linewidth\linewidth\else\Gin@nat@width\fi}
\def\maxheight{\ifdim\Gin@nat@height>\textheight\textheight\else\Gin@nat@height\fi}
\makeatother
\setkeys{Gin}{width=\maxwidth,height=\maxheight,keepaspectratio}
\makeatletter
\def\fps@figure{htbp}
\makeatother

\makeatletter
\@ifpackageloaded{caption}{}{\usepackage{caption}}
\AtBeginDocument{%
\ifdefined\contentsname
  \renewcommand*\contentsname{Table of contents}
\else
  \newcommand\contentsname{Table of contents}
\fi
\ifdefined\listfigurename
  \renewcommand*\listfigurename{List of Figures}
\else
  \newcommand\listfigurename{List of Figures}
\fi
\ifdefined\listtablename
  \renewcommand*\listtablename{List of Tables}
\else
  \newcommand\listtablename{List of Tables}
\fi
\ifdefined\figurename
  \renewcommand*\figurename{Figure}
\else
  \newcommand\figurename{Figure}
\fi
\ifdefined\tablename
  \renewcommand*\tablename{Table}
\else
  \newcommand\tablename{Table}
\fi
}
\@ifpackageloaded{float}{}{\usepackage{float}}
\floatstyle{ruled}
\@ifundefined{c@chapter}{\newfloat{codelisting}{h}{lop}}{\newfloat{codelisting}{h}{lop}[chapter]}
\floatname{codelisting}{Listing}

\makeatother
\makeatletter
\@ifpackageloaded{caption}{}{\usepackage{caption}}
\@ifpackageloaded{subcaption}{}{\usepackage{subcaption}}
\makeatother

\ifLuaTeX
  \usepackage{selnolig}  % disable illegal ligatures
\fi
\usepackage[]{natbib}
\usepackage{bookmark}

\usepackage{url}

\IfFileExists{xurl.sty}{\usepackage{xurl}}{} % add URL line breaks if available
\hypersetup{
  pdftitle={Title},
  pdfauthor={Author 1; Author 2},
  pdfkeywords={3 to 6 keywords, that do not appear in the title},
  colorlinks=true,
  linkcolor={blue},
  filecolor={Maroon},
  citecolor={Blue},
  urlcolor={Blue},
  pdfcreator={LaTeX via pandoc}}

\newcommand{\anon}{1}

\begin{document}

\def\spacingset#1{\renewcommand{\baselinestretch}%
{#1}\small\normalsize} \spacingset{1}

%%%%%%%%%%%%%%%%%%%%%%%%%%%%%%%%%%%%%%%%%%%%%%%%%%%%%%%%%%%%%%%%%%%%%%%%%%%%%%

\if1\anon
{
  \title{\bf One Person, How Many Votes? Demographic Distortions in United States Elections}
  \author{Lee Kennedy-Shaffer\hspace{.2cm}\\
    Department of Biostatistics, Yale School of Public Health}
  \maketitle
} \fi

\if0\anon
{
  \bigskip
  \bigskip
  \bigskip
  \begin{center}
    {\LARGE\bf Title}
\end{center}
  \medskip
} \fi

\bigskip
\begin{abstract}
Representative democracy in the United States relies on election systems that transmit votes into representatives in three key bodies: the two chambers of the federal legislature (House of Representatives and Senate) and the Electoral College, which selects the President and Vice-President. This happens through a process of re-weighting based on geographic units (congressional districts and states) that can introduce substantial distortion. In this paper, I propose quantitative measures of this distortion that can be applied to demographic groups, using Census data, to assess and visualize these distortive effects. These include the absolute weight of votes under these systems and the excess population represented in the bodies through the distortions. Visualizing these metrics from 2000---2020 shows persistent malapportionment in key demographic categories. White (non-Hispanic) residents, residents of rural areas, and owner-occupied households are overrepresented in the Senate and Electoral College; Black and Hispanic people, urban dwellers, and renter-occupied households are underrepresented. For urban residents, this underrepresentation is the equivalent of 25 million fewer residents in the Senate and nearly 5 million in the Electoral College. I discuss implications for further research on the effects of these distortions and their interactions with other features of the electoral system.
\end{abstract}

\noindent%
\vfill

\newpage
\spacingset{1.8} % DON'T change the spacing!

\section{Introduction}\label{sec-intro}

In September 1787, the United States Constitutional Convention finalized decisions that would affect the democratic representativeness of some of the world's most powerful political offices. The proposed Constitution included a bicameral legislature, with representation in the House of Representatives apportioned according to the population of the states and representation in the Senate apportioned to each state itself, and election of the President and Vice-President by the Electoral College (EC), with electors apportioned to each state according to its total number of Representatives and Senators \citep{congressional_research_service_constitution_2025}. From the beginning, the compromise forged between different notions of representation of people (at the time, only white males) and representation of states as political entities \citep[see, e.g.,][]{alexander_representation_2019} created conflict while preserving, for a time, an uneasy union by maintaining the power of both small states and slave-holding states \citep{keyssar_why_2020,dupont_distorting_2024}.

While the system of political representation in these bodies changed substantially over the years, in particular as winner-take-all voting for electors became the norm in the 1830s and as direct election of Senators became required in 1913 \citep{dupont_distorting_2024,congressional_research_service_constitution_2025}, changes to the core apportionment have been few and far between. Only the Reconstruction amendments to the Constitution---which voided the three-fifths clause for representation of enslaved individuals, which eventually increased the power of white Southerners as they continued to deny Black residents their right to vote---and the 23rd Amendment, ratified in 1961---which provided EC representation to the District of Columbia while continuing to deny it House or Senate representation---changed the Constitutional text regarding apportionment \citep{dupont_distorting_2024,congressional_research_service_constitution_2025}.

Two other major changes came in the 1960s, creating the modern form of American democracy. A series of Supreme Court cases beginning with \citet{noauthor_baker_1962} mandated re-drawing of districts in the House of Representatives after every decennial Census and re-apportionment to ensure approximately equal populations of districts within states \citep{baker_reapportionment_1966}. This improved the representativeness of the House, while permitting extreme imbalances in the Senate and EC. Nonetheless, some malapportionment persists in the House due to rounding issues, creeping distortions over time between instances of redistricting, and Constitutional provisions excluding DC and guaranteeing one representative for all states \citep{gaines_apportionment_2009}. In addition, the Voting Rights Act codified political rights regardless of race and improved (but did not perfect) alignment of the voting-eligible and represented populations \citep{berman_minority_2024}.

The effects of these decisions, decades and centuries later, are still being felt, as the malapportionment of representation in the Senate and EC permits distorted and, in some cases, minority rule. The misrepresentation across states is well-documented, although sometimes reinforced by maps by area as described by \citet{wegman_let_2020} and \citet{kennedy-shaffer_land_2021}, tending to surface every two or four years with election cycles. The distorted representation is often expressed in a comparison of the number of residents or voters per senator or elector in big states (such as California) compared to small states (such as Wyoming) \citep[see, e.g.,][]{kallenbach_our_1960, alexander_representation_2019,edwards_why_2004} or by aggregating across states \citep[see, e.g.,][]{cervas_legal_2020}. The consequences of this for unrepresentativeness of these bodies and increased skepticism of the political process are often discussed without disentangling the apportionment itself from other features of the system, such as winner-take-all elections and the possibility of small shifts in vote counts to have an outsized impact on outcomes \citep{dupont_distorting_2024}. Indeed, winner-take-all is often a greater focus than the distortion of the states---especially in inverted presidential elections where the popular vote loser wins the EC---as it is seen to disenfranchise those in each state who vote for the losing candidate \citep[see, e.g.,][]{edwards_why_2004,wegman_let_2020,dupont_distorting_2024}, although malapportionment drove the inversion in 2000.

The consequences of this malapportionment on the relative representation and political power of different demographic groups, rather than states, have been investigated through, for example, comparisons of the demographics of over- and under-represented states \citep{blake_one_2019} or localities \citep{baker_rural_1955}, or the likelihood that a voter from a certain demographic category lives in a tipping-point state \citep{gelman_electoral_2016}. These have shown persistent overrepresentation of white and rural voters \citep{baker_rural_1955,blake_one_2019}. These analyses, however, re-create the core logic of the system: representation partially mediated through states.

Here, I propose statistical metrics to quantify and visualize the distorted representation of demographic groups in the House, Senate, and EC. These measures consider the distorted representations as weighted averages of the true baseline population of the country as a whole, rather than using the states as observations. By using comprehensive state- and district-level Census data, this approach can compare the true population to the represented population, permitting discussion of the over- and under-weighting of certain groups in these bodies. This treats misrepresentation as a constitutional statistical phenomenon that underpins the system as a whole. Additional distortions in the outcomes can arise because of layers of incomplete or inaccurate representation added on top: for example, winner-take-all systems, limitations on voter eligibility, and low voter turnout. Analysis of these can be added to this foundation with further considerations of counterfactuals under different electoral systems and voting patterns. The results presented here, however, require only a comparison of the actual distortion induced by the allocations within these three bodies. These metrics, and the demographic variables considered, are explained in Section~\ref{sec-meth}. The results and visualizations are presented in Section~\ref{sec-res} and in the accompanying R Shiny application (\url{https://bit.ly/Elec-Weights}). Finally, the implications of these results and the advantages and limitations of this approach, are discussed in Section~\ref{sec-disc}.

\section{Methods}\label{sec-meth}

\subsection{Census Data}

Data from the U.S. Census Bureau API were accessed using a developer key \citep{united_states_census_bureau_developers_2025} with the tidycensus (v1.7.3) package in R \citep{walker_analyzing_2023,walker_tidycensus_2025}. Short-form Census data were acquired from the 2000, 2010, and 2020 Censuses using Summary File 1 for 2000 and 2010 and the Demographic and Housing Characteristics summary file for 2020. The 2010 and 2020 data are aggregated by congressional district and the 2000 data by state. The specified year represents the year in which the Census data collection and the election occurred; congressional districts (and thus electoral votes) are later re-apportioned according to those populations, but this has not occurred by that year's election. For example, the 2020 results would use 2020 Census population numbers but the districts in place for the 2020 elections, prior to re-apportionment. Because of this, districts in a Census year are at their peak divergence from the current population.

Demographic characteristics gathered include the following:

\begin{itemize}
\tightlist
\item Race/Ethnicity: categorized into Hispanic or Latino ethnicity of any race and, among those not reporting Hispanic or Latino ethnicity, either the single race reported (White, Black, Asian, American Indian or Alaska Native (AIAN), Native Hawaiian or Pacific Islander (NHOPI), Other race) or a category for anyone reporting multiple races.
\item Age Category: categorized into ages 0--17, 18--39, 40--64, or 65+ years old.
\item Sex: only male or female are reported.
\item Rural/Urban Status: categorized into rural or urban (2020) or rural, urban cluster, or urbanized area (2000 and 2010). For trend comparisons, urban cluster and urbanized area are combined into a single urban category. For the 2020 Census, ``rural areas comprise open country and settlements with fewer than 2,000 housing units and 5,000 residents''; all other areas are considered urban \citep{sanders_rural_2025}. For previous years, urban areas were split into urbanized areas with 50,000 or more people and urban clusters with 2,500 to 49,999 people \citep{sanders_rural_2025}.
\item Housing Status: categorized into renter-occupied, owner-occupied with a mortgage, and owner-occupied free and clear (2010 and 2020) or renter-occupied or owner-occupied (2000). For trend comparisons, the two owner-occupied subcategories are combined into a single owner-occupied category. Housing status is counted by households, rather than population, so all comparisons and weights are calculated with the household unit of analysis.
\end{itemize}

These characteristics have been selected for their accessibility and (relatively) consistent definitions across recent decennial Censuses, as well as their potentially important roles in determining preferred policies among U.S. residents. Certainly, other demographic variables could be important as well. R code to conduct similar analyses with other variables is available at the GitHub repository (\url{https://bit.ly/Elec-Analysis}). 

\subsection{Baseline Population}

For baseline populations to calculate the proportion of the overall population falling into different demographic categories, the population in the U.S. was summed. Baseline populations include all residents counted in the decennial Census, as that is the population used for apportionment of the U.S. Congress \citep{congressional_research_service_constitution_2025}. Note that this is distinct in several ways from the population eligible to vote in federal elections. Major restrictions on the franchise that may shift the population demographics include the requirement to be eighteen years old \citep{congressional_research_service_constitution_2025}, rescinded voting rights for those with felony convictions in some states \citep{uggen_locked_2024}, and the long-standing (but not permanent) restriction of the franchise to citizens \citep{kennedy_voters_2022}.

Primary analyses use a baseline population that includes the population of the District of Columbia, but do not include the populations of Puerto Rico or other U.S. territories. This is to maintain consistency in the baseline population, as the District of Columbia does have three Electoral College votes, but none of the territories have Electoral College votes \citep{congressional_research_service_constitution_2025}. Neither the District of Columbia nor any of the territories have voting representation in the House of Representatives or Senate. Alternative analyses using a baseline population that includes the territories, or that excludes the District of Columbia, are presented in the interactive R Shiny application (\url{https://bit.ly/Elec-Weights}).

\subsection{Metrics for Quantification and Visualization}

Four metrics are used to quantify and visualize the relative representation of demographic categories: proportion of the represented population, absolute weight, relative weight, and excess population. Each is computed for the different demographic variables and for each of the three representative bodies (House of Representatives, Senate, and Electoral College).

All metrics rely on the proportion of the body represented by each relevant unit for that body: congressional district for the House of Representatives, state for the Senate, and elector-apportioning unit for the Electoral College. There are 435 total votes in the House of Representatives, with each district allocated one vote. There are 100 total votes in the Senate, with each state allocated two votes. In the current Electoral College, there are 538 total votes; the elector-apportioning unit is the District of Columbia (3 votes), the five congressional districts of Maine and Nebraska (1 vote each), the state of Maine and Nebraska as a whole (2 votes each), and the remaining 48 states (with a number of votes equal to the number of congressional districts in the state plus two). Note that for the Electoral College, the unit depends on the method of awarding electors from a state; this has a minor difference on the outcome, however, as awarding of electors by district does not lead to substantial additional misrepresentation since congressional districts have roughly equal population as of the prior Census. So a shift by more states to district-based elector awarding would not substantially change these numbers, which is related to the constant expectation of votes awarded noted in \citet{thomas_estimating_2013}.

The represented population proportion for a demographic category for a body is determined by computing a weighted sum of the proportion of the population of each relevant unit in that category, weighted by the proportion of the body's total votes apportioned to the unit. For comparison, the proportion of the baseline population in that demographic category is given as well. Letting $b=0,1,2,3$ represent the baseline and the three bodies considered (House, Senate, Electoral College), respectively, $u=1,\ldots,U_b$ the relevant units for body $b$, $\pi_{b,u}$ the proportion of the population in unit $u$ in body $b$ that is in the demographic category of interest, and $v_{b,u}$ the number of votes in body $b$ apportioned to unit $u$ (or, for $b=0$, the total population in unit $u$), the represented population proportion for that category in that body $\pi_b$ is given by:

$$
\pi_b = \frac{\sum_{u=1}^{U_b} \pi_{b,u} v_{b,u}}{\sum_{u=1}^{U_b} v_{b,u}}.
$$

The absolute weight for a demographic category for a body is the category's represented population proportion in that body divided by the category's proportion of the baseline population. For bodies $b=1,2,3$ (House, Senate, Electoral College, respectively), the absolute weight $w_b$ for a demographic category of interest is given by:

$$
w_b = \frac{\pi_b}{\pi_0}.
$$

The relative weight for a demographic category for a body is the absolute weight for that category in that body divided by the absolute weight for the referent category for that body. The referent categories are: white, aged 0--17, female, rural, and renter-occupied.

An absolute weight greater than one indicates a demographic category whose represented population proportion is greater than its baseline population proportion. A relative weight greater than one indicates a demographic category whose absolute weight is greater than that of the referent category (i.e., it is more over-represented than the referent category).

The excess population for a demographic category for body $b$ is the product of the total baseline population ($\sum_{u=1}^{U_0} v_{0,u}$) and $\pi_0 - \pi_b$, the difference between the baseline population proportion and the represented population proportion for that category. This represents the change in population that demographic category would need, in the fixed baseline population, for its population proportion to equal its \emph{represented} population proportion. For any demographic variable, the excess population across all categories sums to zero. A demographic category with an absolute weight greater than one will have a positive excess population; a category with an absolute weight less than one will have a negative excess population.

The calculations are conducted for each of three Census years: 2000, 2010, and 2020.

\section{Results}\label{sec-res}

\begin{figure}

\centering{

\includegraphics[width=\textwidth]{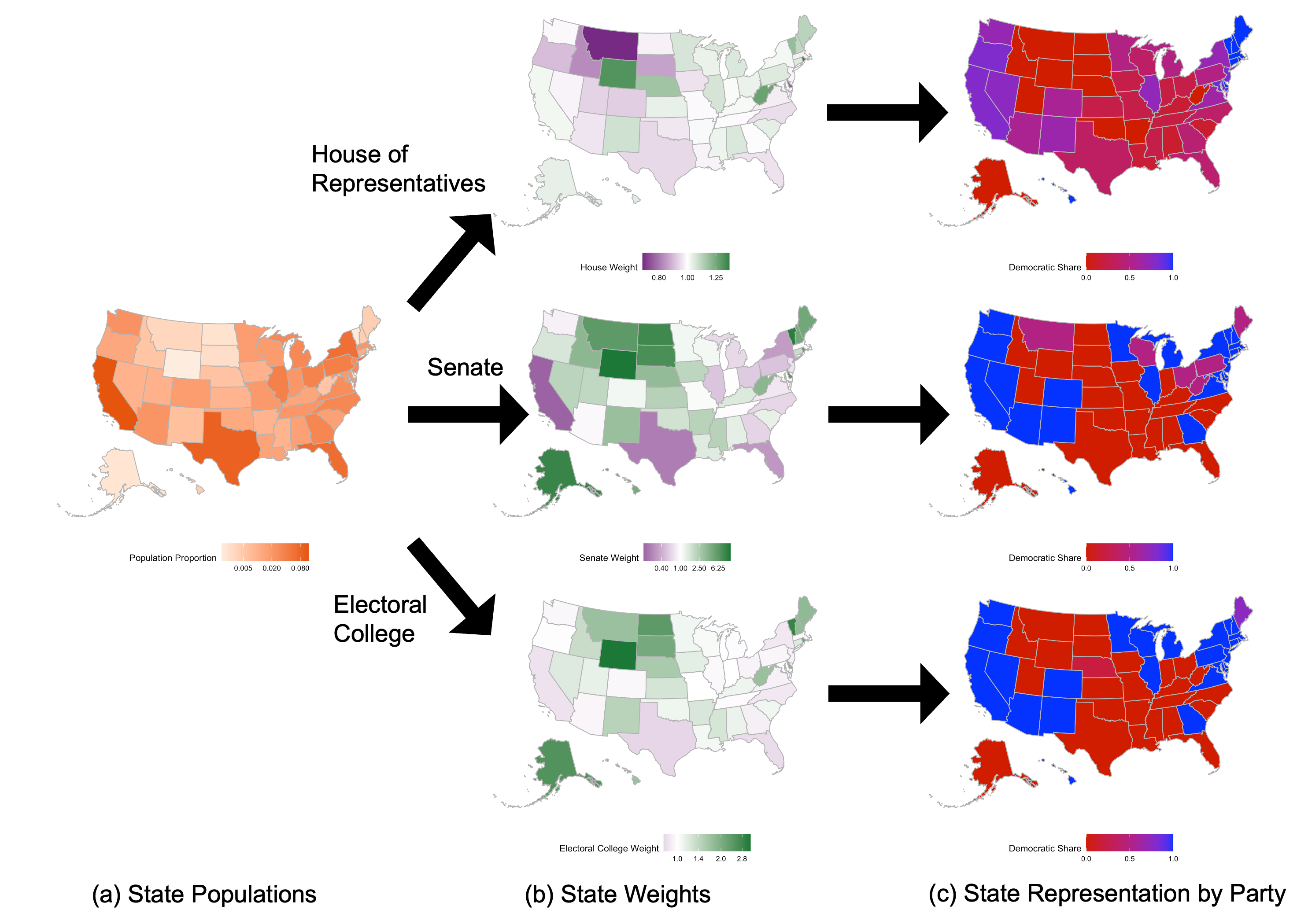}

}

\caption{\label{fig-schematic}Population proportions (a), absolute population weights in each body (b), and Democratic representation share in the 2020 election for each body (c) for each state. Democratic representation share shows the proportion of House (top row) and Senate (middle row) voting representatives for the state caucusing with the Democrats (including independents) or the proportion of electors (bottom row) pledged to the Democratic ticket. The first set of arrows is the result of apportionment of seats and the relative populations, while the second set of arrows is the result of voter eligibility, registration, turnout, and decisions, and the allocation of representation based on votes (i.e., winner-take-all for most contests).}

\end{figure}

The representation weight for each state's population for each body is shown in Figure~\ref{fig-schematic}, along with the eventual 2020 representation by party for each state in each body \citep[using election results from][]{mit_election_data_and_science_lab_us_2017,mit_election_data_and_science_lab_us_2017-1,mit_election_data_and_science_lab_us_2017-2}. These state (and within-state district) weights are what determine the differences between the baseline population proportions and represented proportions for each demographic category. The Senate and Electoral College demonstrate the well-known pattern of underrepresentation of large states and overrepresentation of small states, while the House has a less clear picture, as its malapportionment relies largely on demographic trends since the last reapportionment and on the allocation algorithm leading to major differences in representation among small states (for example, compare the underrepresented Montana to the overrepresented Rhode Island and Wyoming).

\subsection{Represented Proportions and Weights: 2020}

\begin{figure}

\centering{
\begin{subfigure}{0.45\textwidth}
	\caption{By Race/Ethnicity}
	\includegraphics[width=\textwidth]{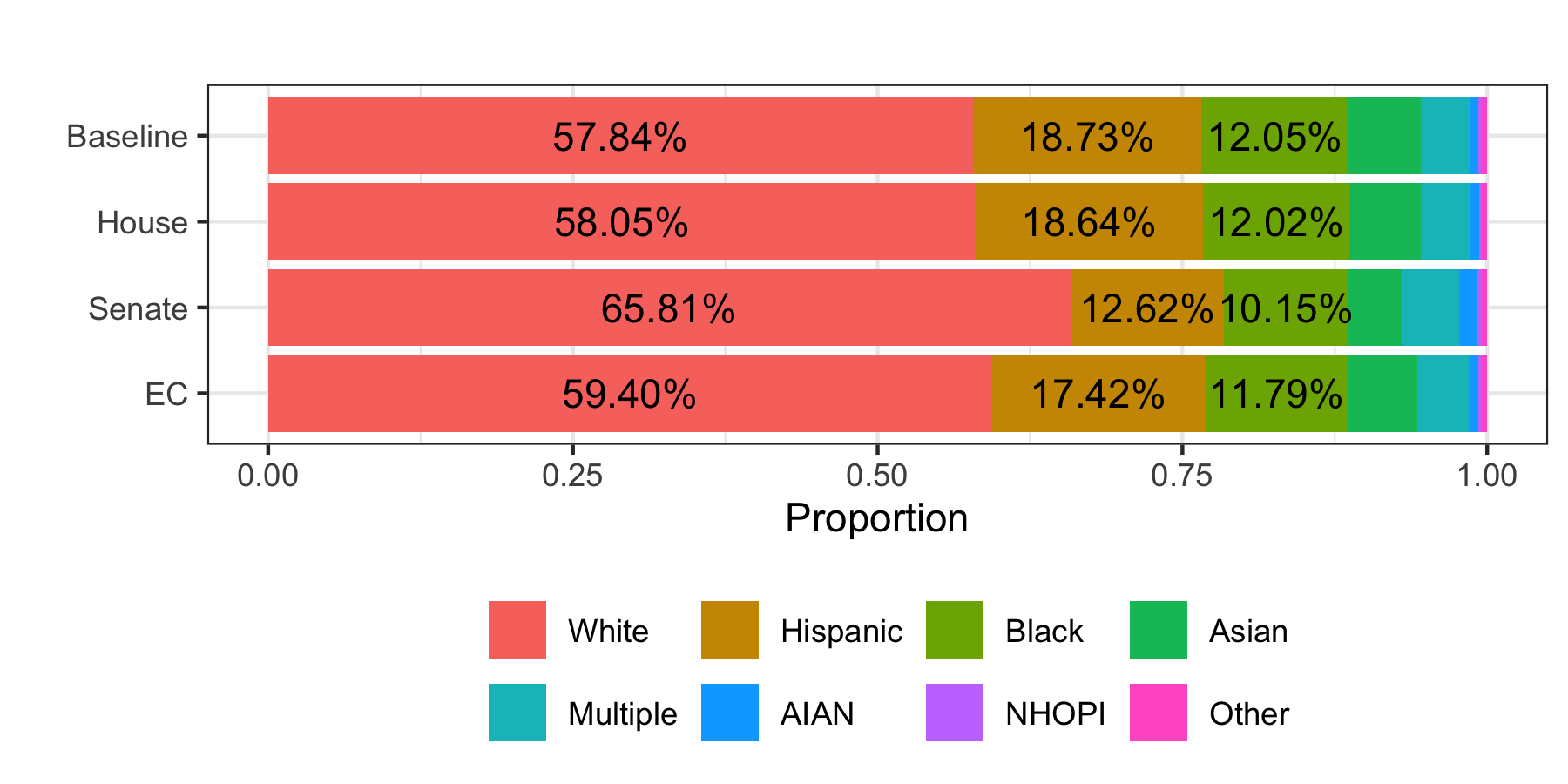}
\end{subfigure}
\begin{subfigure}{0.45\textwidth}
	\caption{By Age Category}
	\includegraphics[width=\textwidth]{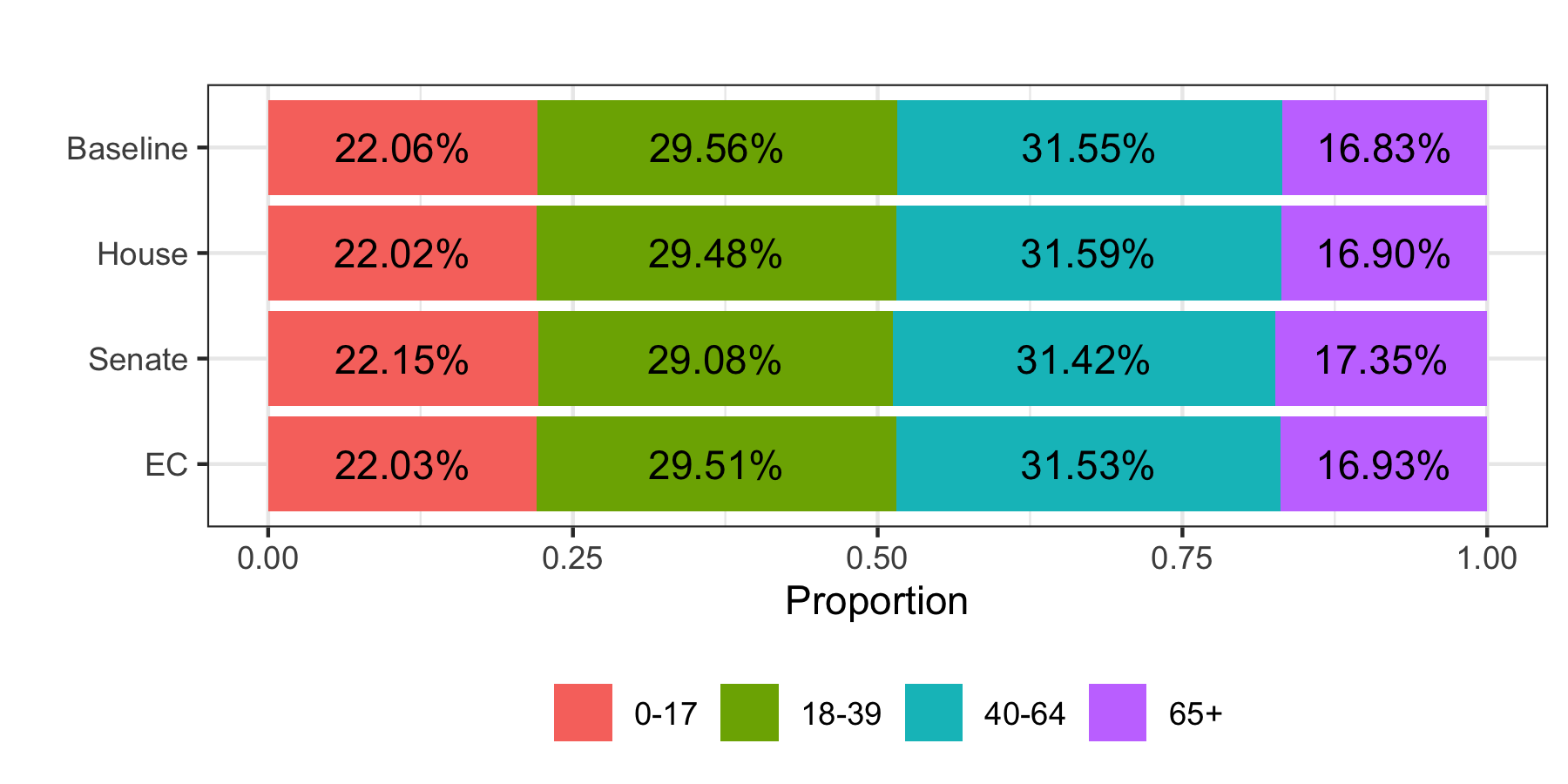}
\end{subfigure}
\begin{subfigure}{0.45\textwidth}
	\caption{By Sex}
	\includegraphics[width=\textwidth]{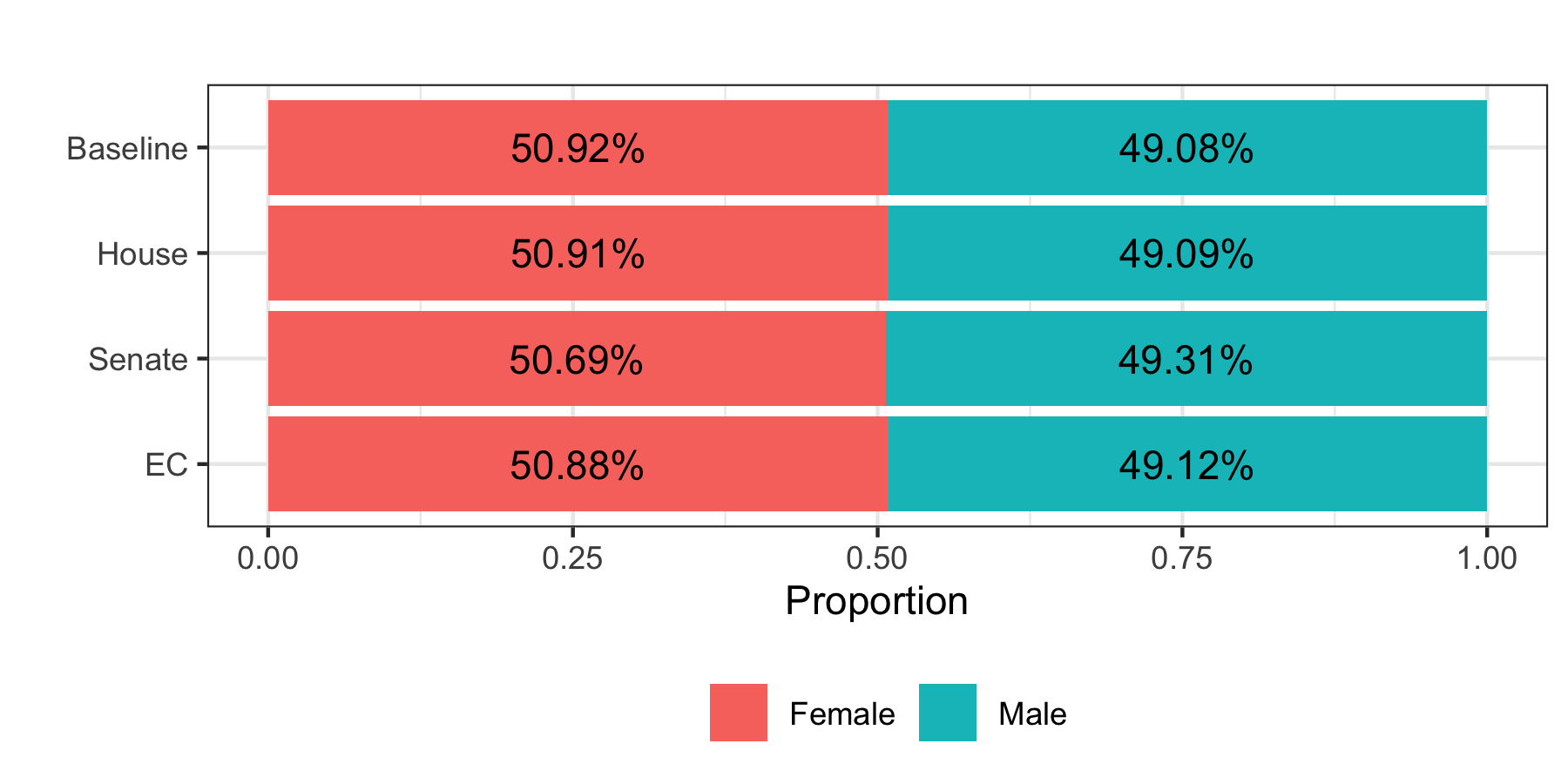}
\end{subfigure}
\begin{subfigure}{0.45\textwidth}
	\caption{By Rural/Urban Status}
	\includegraphics[width=\textwidth]{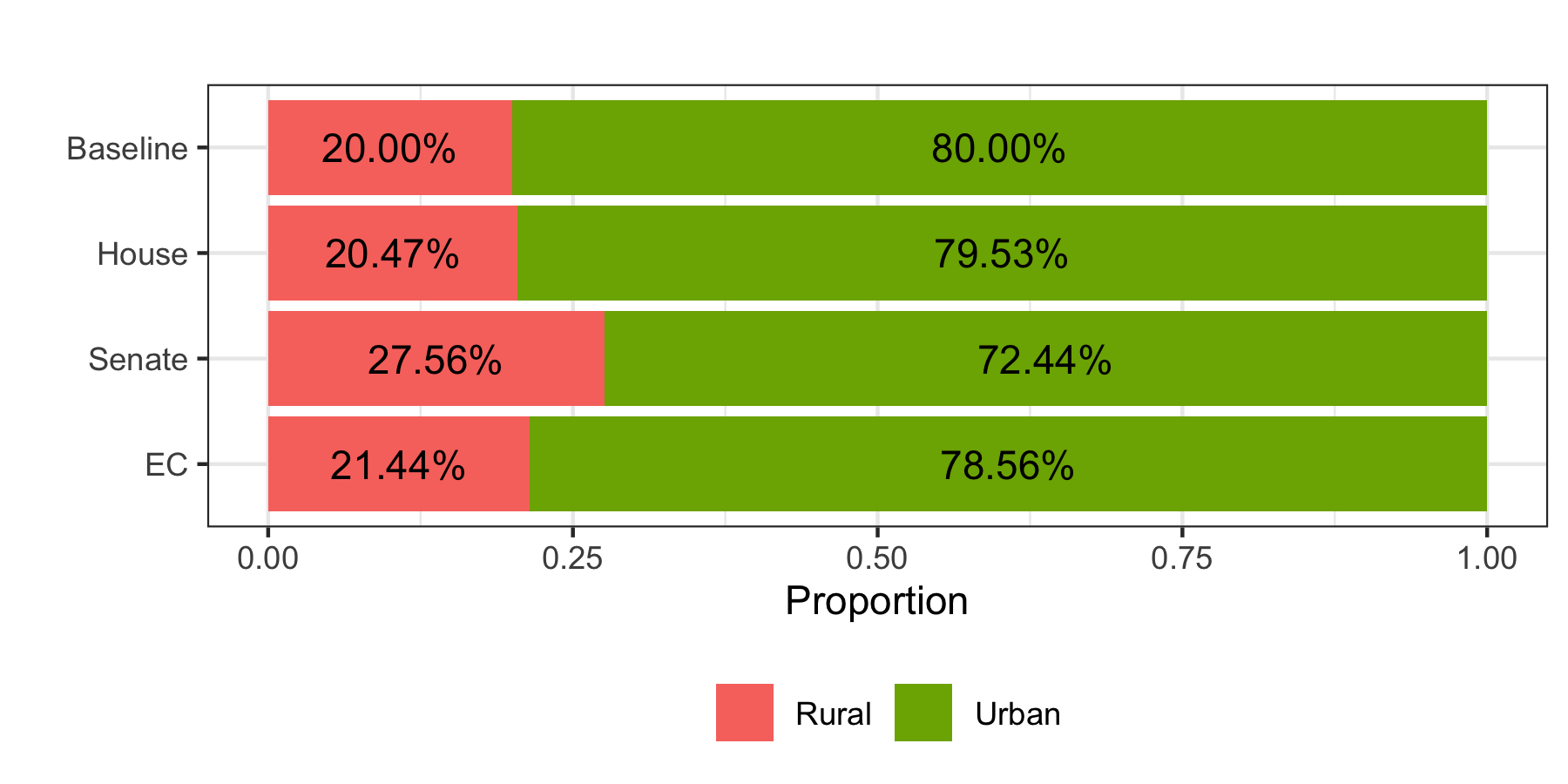}
\end{subfigure}
\begin{subfigure}{0.45\textwidth}
	\caption{By Housing Status}
	\includegraphics[width=\textwidth]{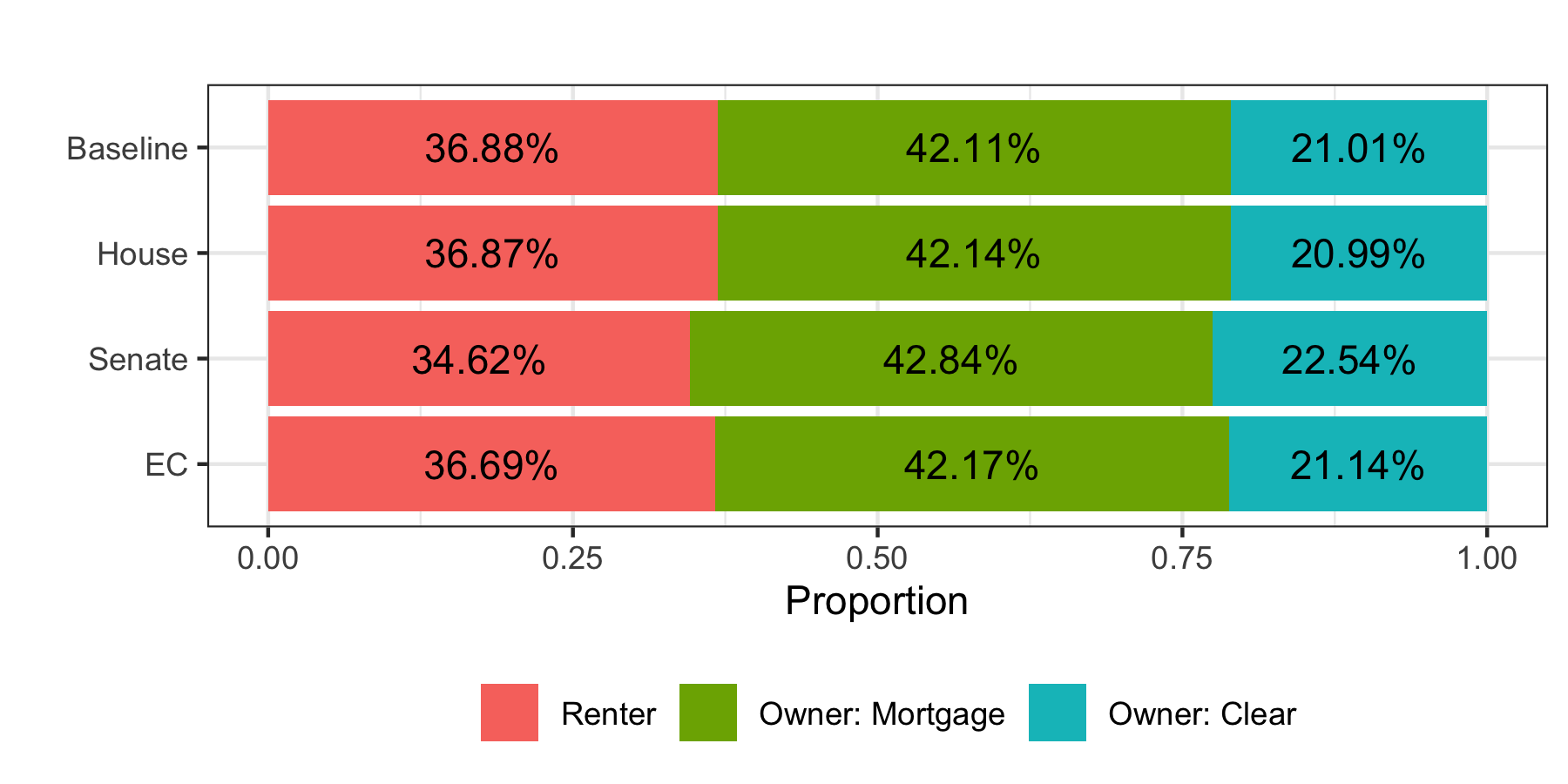}
\end{subfigure}
}

\caption{\label{fig-proportions}Baseline and represented population proportions for each body, by demographic variable, as of the 2020 Census. Proportions under 10\% are not labelled. Note: for housing status (e), proportions are of the total households instead of total population.}

\end{figure}

Figure~\ref{fig-proportions} shows the baseline and represented population proportions by race/ethnicity, age category, sex, rural/urban status, and housing status as of the 2020 Census. For all demographic variables, the House of Representatives represented population is fairly similar to the baseline population. The Electoral College shows large differences, particularly for race/ethnicity, rural/urban status, and housing status, and the Senate shows the largest differences for those three variables. Age category and sex generally exhibit fairly stable representation.

The extent of this distorted representation is seen more clearly in the absolute weights and excess population for each demographic category, as well as the relative weights compared to the referent category for each variable; these are displayed in Table~\ref{tbl-wts}. On sex and age category, the distortions are relatively minor, with no absolute weights greater than 1.031 or less than 0.984.

On race and ethnicity, this demonstrates overrepresentation of the non-Hispanic white population and corresponding underrepresentation of the Hispanic, non-Hispanic Black, and non-Hispanic Asian populations. This distorted representation is equivalent to 26 million more white voters in the Senate and over 5 million in the Electoral College, with the represented Hispanic population missing 20 million voters in the Senate and over 4 million in the Electoral College. Smaller racial categories---specifically the American Indian and Alaska Native and Native Hawaiian and Other Pacific Islander groups---are overrepresented due to their concentration in relatively few small states. 

\spacingset{1}
{\small
\begin{longtable}[]{@{}l|lll|lll|lll|@{}}
\caption{{\small Absolute weight (AW), relative weight (RW), and excess population (EP) for each body, by demographic variable, as of the 2020 Census.}}\label{tbl-wts}\tabularnewline
\toprule\noalign{}
& \multicolumn{3}{c|}{House} & \multicolumn{3}{c|}{Senate} & \multicolumn{3}{c|}{Electoral College} \\
Category & AW & RW& EP & AW & RW& EP & AW & RW& EP \\
\midrule\noalign{}
\endfirsthead
\toprule\noalign{}
& \multicolumn{3}{c|}{House} & \multicolumn{3}{c|}{Senate} & \multicolumn{3}{c|}{Electoral College} \\
Category & AW & RW& EP & AW & RW& EP & AW & RW& EP \\
\midrule\noalign{}
\endhead
\bottomrule\noalign{}
\endlastfoot
\multicolumn{10}{l|}{Race/Ethnicity} \\
White & 1.004 & 1.000 &    719,279 & 1.138 & 1.000 &  26,430,619 & 1.027 & 1.000 &   5,174,736 \\ 
  Hispanic & 0.995 & 0.991 &   -314,020 & 0.674 & 0.592 & -20,254,385 & 0.930 & 0.906 &  -4,337,322 \\ 
  Black & 0.997 & 0.993 &   -112,187 & 0.842 & 0.740 &  -6,307,179 & 0.978 & 0.953 &   -860,494 \\ 
  Asian & 0.989 & 0.985 &   -213,231 & 0.753 & 0.661 &  -4,854,679 & 0.956 & 0.931 &   -863,042 \\ 
  Multiple & 0.996 & 0.992 &    -52,668 & 1.134 & 0.997 &   1,815,331 & 1.025 & 0.998 &    336,041 \\ 
  AIAN & 0.995 & 0.992 &    -10,626 & 2.208 & 1.940 &   2,719,582 & 1.208 & 1.177 &    469,061 \\ 
  NHOPI & 0.995 & 0.991 &     -3,048 & 1.991 & 1.750 &    616,498 & 1.182 & 1.151 &    113,318 \\ 
  Other & 0.992 & 0.988 &    -13,499 & 0.902 & 0.793 &   -165,785 & 0.981 & 0.955 &    -32,298 \\ 
  \midrule\noalign{}
  \multicolumn{10}{l|}{Age Category} \\
  0-17 & 0.999 & 1.000 &  -107,823 & 1.004 & 1.000 &   300,970 & 0.999 & 1.000 &   -86,058 \\ 
  18-39 & 0.998 & 0.999 &  -238,100 & 0.984 & 0.980 & -1,592,060 & 0.998 & 1.000 &  -148,594 \\ 
  40-64 & 1.001 & 1.003 &   109,005 & 0.996 & 0.992 &  -436,277 & 0.999 & 1.000 &   -90,305 \\ 
  65+ & 1.004 & 1.006 &   236,918 & 1.031 & 1.027 &  1,727,367 & 1.006 & 1.007 &   324,957 \\ 
  \midrule\noalign{}
  \multicolumn{10}{l|}{Sex} \\
  Female & 1.000 & 1.000 &  -25,143 & 0.995 & 1.000 & -760,319 & 0.999 & 1.000 & -106,998 \\ 
  Male & 1.000 & 1.000 &   25,143 & 1.005 & 1.009 &  760,319 & 1.001 & 1.001 &  106,998 \\ 
  \midrule\noalign{}
  \multicolumn{10}{l|}{Rural/Urban Status} \\
  Rural & 1.023 & 1.000 &   1,542,942 & 1.378 & 1.000 &  25,058,678 & 1.072 & 1.000 &   4,776,461 \\ 
  Urban & 0.994 & 0.972 &  -1,542,942 & 0.905 & 0.657 & -25,058,678 & 0.982 & 0.916 &  -4,776,461 \\ 
  \midrule\noalign{}
  \multicolumn{10}{l|}{Housing Status (Household Level)} \\
  Renter & 1.000 & 1.000 &    -8,590 & 0.939 & 1.000 & -2,861,409 & 0.995 & 1.000 &  -239,505 \\ 
  Owner: Mortgage & 1.001 & 1.001 &    37,610 & 1.017 & 1.084 &   919,431 & 1.001 & 1.007 &    73,647 \\ 
  Owner: Clear & 0.999 & 0.999 &   -29,020 & 1.073 & 1.143 &  1,941,978 & 1.006 & 1.011 &   165,857 \\ 
%  Owner & 1.000 & 1.000 &     8,590 & 1.036 & 1.103 &  2,861,409 & 1.003 & 1.008 &   239,505 \\ 
\end{longtable}
}
\spacingset{1.8}

Geographically, rural populations are consistently overrepresented across the three bodies. An urban resident has 0.972, 0.657, and 0.916 times as much representation as a rural resident in the House, Senate, and Electoral College, respectively. The scale of the excess population for rural residents is similar to that for non-Hispanic white residents: around 1 million in the House, 25 million in the Senate, and 5 million in the Electoral College. Housing status shows a less consistent pattern across the three bodies: households owning their homes free and clear are underrepresented in the House, but overrepresented in the Senate (absolute weight: 1.143) and the Electoral College (absolute weight: 1.006); renters are underrepresented throughout.

\subsection{Trends Over Time}

\begin{figure}

\centering{
\begin{subfigure}{0.3\textwidth}
	\caption{Race/Ethnicity}
	\includegraphics[width=\textwidth]{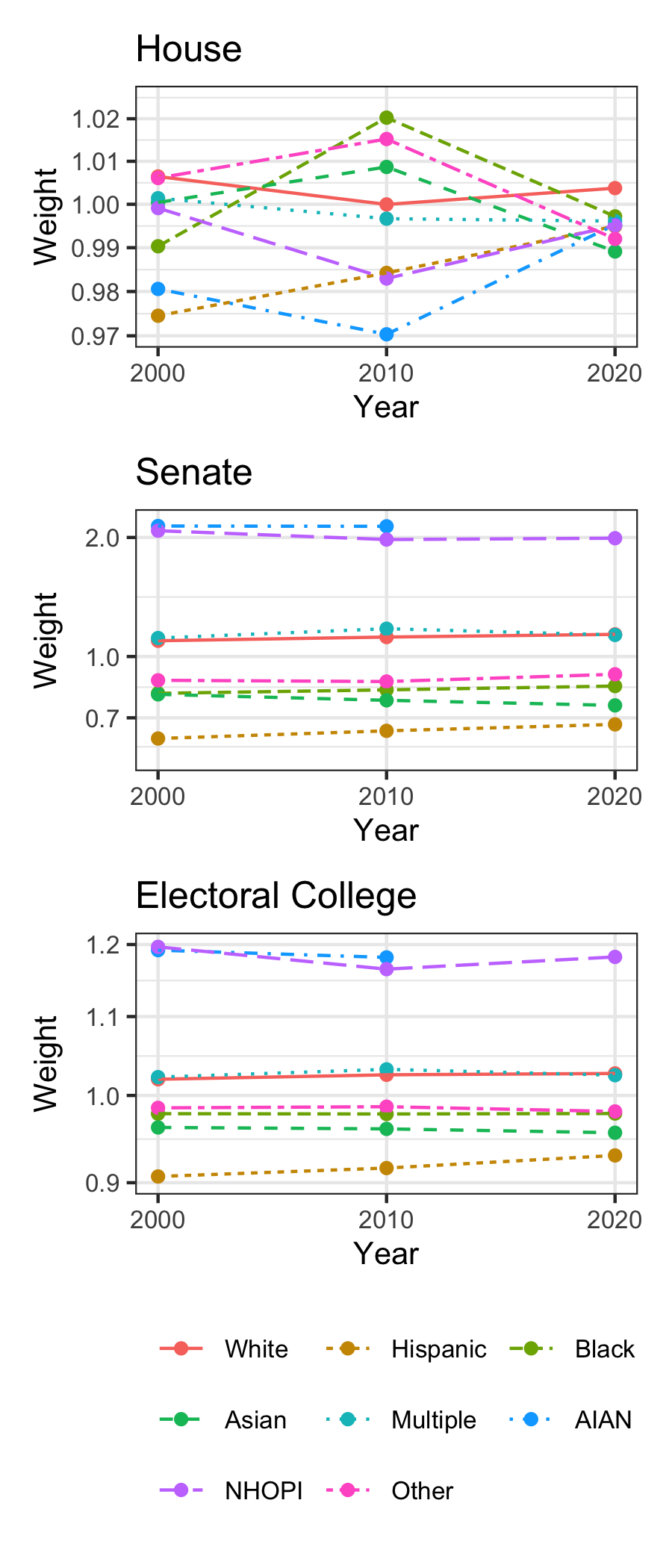}
\end{subfigure}
\begin{subfigure}{0.3\textwidth}
	\caption{Rural/Urban Status}
	\includegraphics[width=\textwidth]{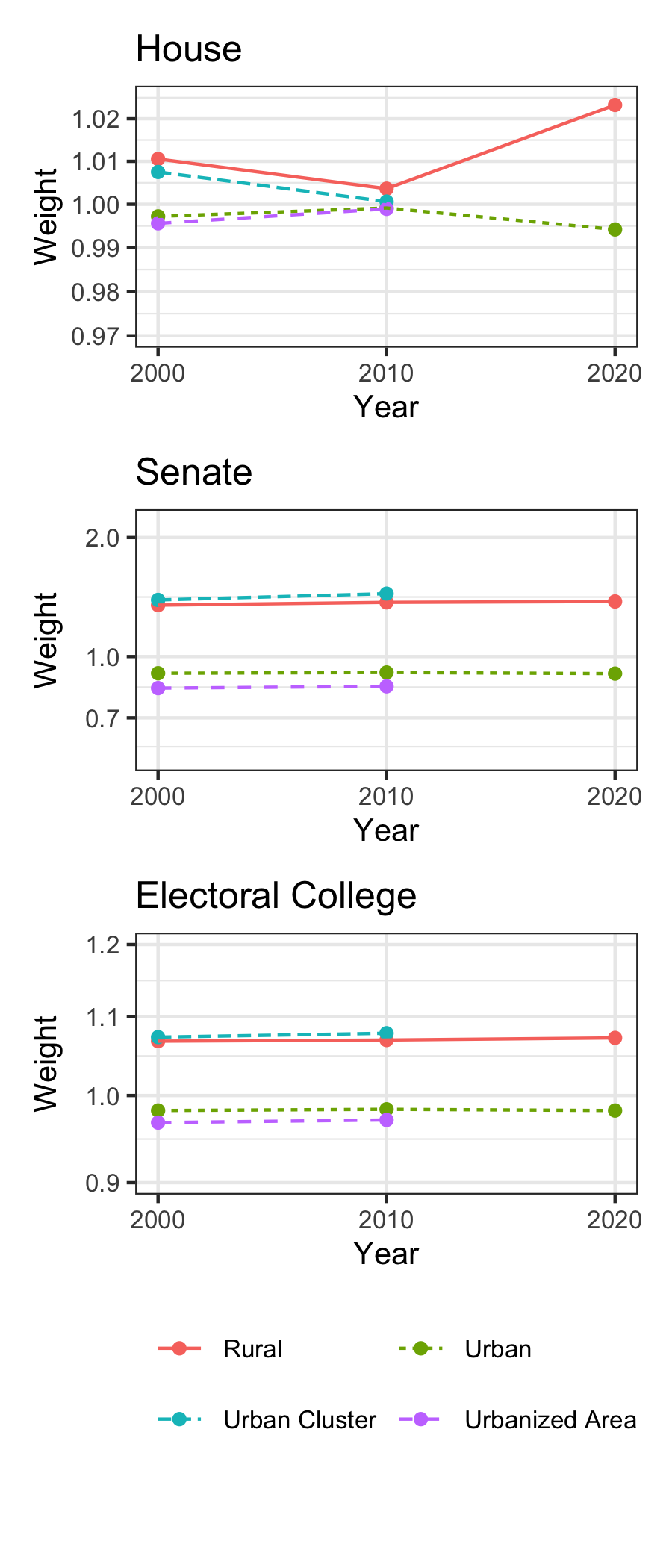}
\end{subfigure}
\begin{subfigure}{0.3\textwidth}
	\caption{Housing Status}
	\includegraphics[width=\textwidth]{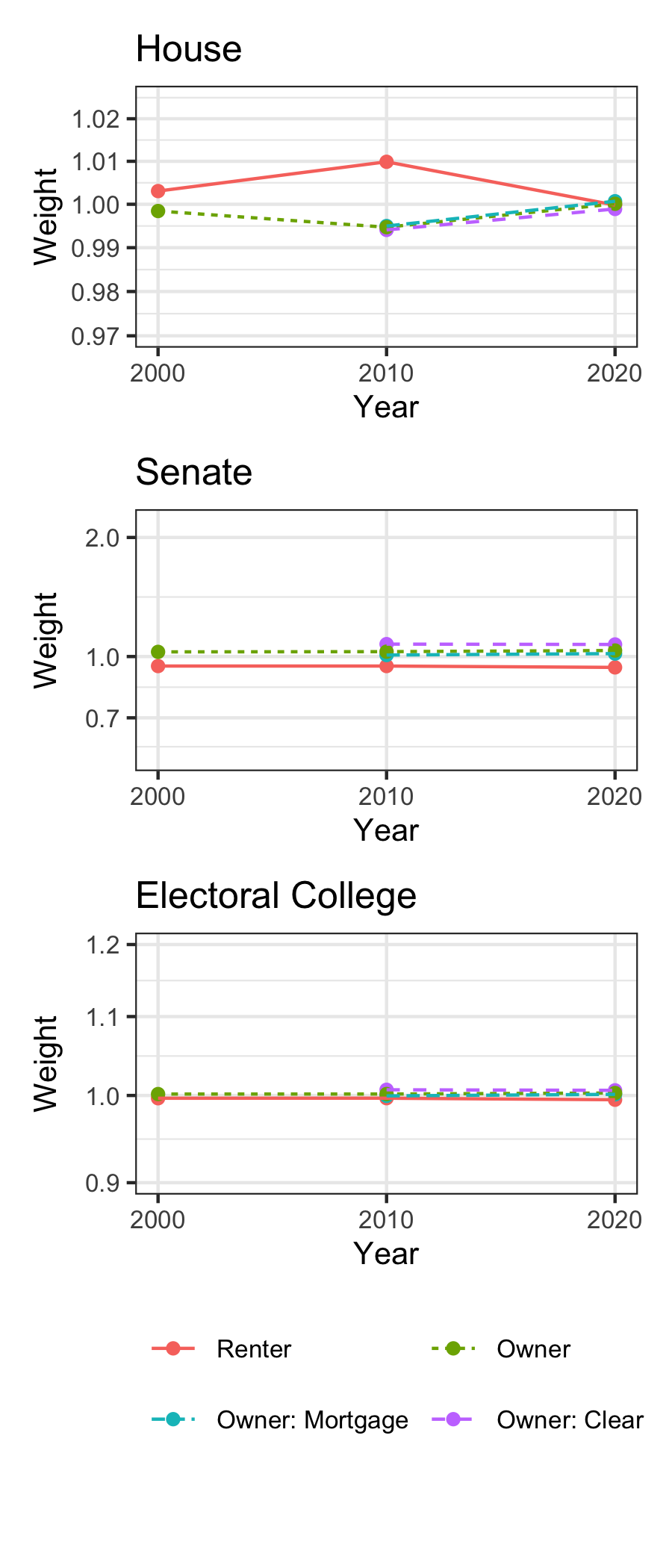}
\end{subfigure}
}
\caption{\label{fig-trends}Trends in absolute weights for each body---House, Senate, and Electoral College---by race/ethnicity, rural/urban status, and housing status, as of the 2000, 2010, and 2020 Censuses. Notes: For rural/urban status (b), urban cluster and urbanized area categories were not used in the 2020 Census; for comparisons, those categories are combined into a single urban category in 2000 and 2010. For housing status (c), weights are based on total households instead of total population. Owner-occupied status was not separated into owner-occupied with a mortgage vs. free and clear ownership in the 2000 Census; for comparisons, those categories are combined into a single owner category in 2010 and 2020.}

\end{figure}

Trends in the absolute weights for the three demographic groups with notable discrepancies (race/ethnicity, rural/urban status, and housing status) across the three most recent Censuses (2000, 2010, and 2020) are shown in Figure~\ref{fig-trends}. All three exhibit similar patterns: high variability in weights across years for the House (including some categories that change between underrepresented and overrepresented), but stability in the ordering of categories and approximate weights across years in the Senate and Electoral College. Some small trends can be observed, however. For one, Hispanic residents have become less underrepresented over time, while Asian residents have become more underrepresented. Renters have become more underrepresented among households, mostly contrasted with increasing overrepresentation of owners with a mortgage.

\subsection{Additional Results}

Additional results, including figures and tables, are available for the demographic variables sex and age category, and for the three Census years included, as well as for alternative specifications of the baseline population (i.e., excluding the District of Columbia or including Puerto Rico). These can be accessed through the GitHub repository (\url{https://bit.ly/Elec-Analysis}) or the interactive R Shiny application (\url{https://bit.ly/Elec-Weights}).

\section{Discussion}\label{sec-disc}

The metrics and visualizations presented here allow for a straightforward consideration of the distortive effects of malapportionment in United States legislative and executive elections. Overrepresentation of non-Hispanic white residents, those living in rural areas, and owner-occupied households in the Senate and, to a lesser extent, the Electoral College is clearly demonstrated. This patterns remain remarkably stable over the past three decennial Censuses (i.e, from 2000 to 2020), covering six presidential and eleven congressional elections.

In some cases, these results align with those from other approaches. On racial demographics, the Electoral College has been found to favor states with whiter populations \citep{blake_one_2019}, overweight white votes \citep{merling_electoral_2016}, and create situations where white voters are more likely to tip the election \citet{gelman_electoral_2016,graphic_detail_how_2020}. On geographic representation, the various bodies of the U.S. federal government have long been known to have a bias toward rural Americans \citep[e.g.,][]{baker_reapportionment_1966}. Although the reapportionment cases in the 1960s largely remedied this issue in the House of Representatives as a whole \citep{eisler_justice_1993}, it remained in the Senate and Electoral College and has persisted to the present day, with attendant policy consequences \citep{badger_as_2016}. Indeed, the racial and rural biases are not separable, as non-white Americans disproportionately live in urban areas; this connection was noted as far back as \citet{noauthor_dyer_1956} and \citet{baker_reapportionment_1966}.

This contravenes some alternative approaches as well. For example, \citet{melotte_analysis_2024} analyzed rural voter power and concluded that it was not disproportionate in the Senate or EC. Specifically, the author determined that overrepresented states were not more rural than average. By using the state as the unit, however, that analysis replicates the basic structure of the Senate and EC, which is responsible for the inequality. Indeed, there are both overrepresented predominantly-rural states (e.g., Vermont and Wyoming) and predominantly-urban states/units (e.g., Rhode Island and DC). More importantly for the population distortion, however, California, New Jersey, and Florida, for example, are underrepresented and overwhelmingly urban. By using the population as a whole as a baseline, then, the actual distortion becomes more apparent.

Nonetheless, there are questions raised by these differences. For example, the measure used to distinguish rural vs. urban areas may be important \citep{sanders_rural_2025}. In addition, how that translates into actual power in Senate votes or the selection of the president depends on more than just the distortion of representation.

\subsection{Advantages and Limitations of this Analysis Approach}
This highlights the key trade-offs of these metrics compared to those used elsewhere. The simplicity of the analysis and interpretation provides one advantage, as does putting differing demographic characteristics on comparable scales (whether absolute weights or excess population). This also provides for ease of visualization, which can be important for conveying information broadly.

Another advantage is the stability of this approach: as seen in the results, these weights persist across decades and elections. In contrast, analysis of battleground or tipping-point states and their characteristics depend specifically on the parties' and candidates' relative strengths in different states, which change by election \citep{burmila_electoral_2009}. Indeed, the defining battleground states of the 2000 and 2004 presidential elections (Florida and Ohio) have since become reliably Republican, while new battleground states emerged in the 2016--2024 elections (e.g., Wisconsin, Michigan, Georgia, Arizona).

This brings a key drawback, however, as the distortions identified here are more distal to election and policy outcomes than in some other methods. In particular, the weights presented here are distinct from notions of ``actual voters represented'' as defined in \citet{kallenbach_our_1960} (how many votes in the election are embodied in each electoral college voter) and of ``voting power'' as defined in, among others, \citet{banzhaf_one_1968} and \citet{grofman_thinking_2005} (which incorporates the probability any one voter has of tipping the election in their candidate's favor under a model of voting patterns within each voting unit). The latter is itself not without controversy; see \citet{sickels_power_1969} for a contemporary rebuttal. Related ideas are used to analyze the demographic distortions of the Electoral College by, e.g., \citet{gelman_electoral_2016} and \citet{wang_best_2020}. More generally, the path from apportionment to election outcome (and, even further, policy outcome) is a lengthy one, affected, as previously mentioned, by the candidates and parties \citep{burmila_electoral_2009}, voter eligibility \citep{kennedy_voters_2022,uggen_locked_2024}, registration, and turnout \citep{barber_400_2022,berman_minority_2024}, voting preferences on election day, winner-take-all voting systems \citep{hoffman_illegitimate_1996,thomas_estimating_2013}, and other factors.

Those alternative approaches rely on models (explicit or implicit) for voter behavior under the system and (generally untestable) assumptions about counterfactuals in alternative voting systems. See \citet{thomas_estimating_2013} for one example of an explicit model. These provide additional explanatory power of the results drawn from those assumptions and models, but at the cost of additional assumptions. In particular, voters may act in relation to their perceived voting power, which may change overall voting behavior if the system is changed. In contrast, this study makes no particular claims about the impact of the distortion on elections, thus avoiding the need for counterfactual scenarios. Indeed, the metrics presented here may provide a model-free foundation on which to layer model-based analyses that can further illuminate this issue and clarify the differing assumptions that underpin conflicting results from previous model-based approaches \citep[see]{longley_biases_1992}.

A related note of caution for this approach (which is relevant to other demographic analyses as well) is that the demographic categories presented here are hardly monolithic political entities. They do not necessarily exercise political power as a bloc, have homogeneous policy preferences, or even identify as an identity group with cohesive interests. Because of this, overrepresentation here does not necessarily indicate greater political power. Moreover, overrepresented groups that are nonetheless a substantial minority in every election unit may have limited ability to exercise their overrepresentation and remain underrepresented in winner-take-all elections. For example, the overrepresentation of American Indian/Alaska Native and Native Hawaiian and Other Pacific Islander people occurs because of relatively high concentrations (although not majorities or even pluralities) in a few small states, such as Alaska, Hawaii, and Oklahoma.  This does not necessarily translate to electing the preferred representatives of those groups in those states, nor would it necessarily confer those representatives with sufficient power to advocate for the preferred policies of people in those groups. This difference is exacerbated if voting restrictions are more burdensome in certain groups as well, as that can magnify underrepresentation or counteract overrepresentation.

\subsection{Data Limitations}

Additional limitations arise from the data used. For one, longitudinal comparisons are challenging as Census questions and methodologies may have changed in ways that affect responses. I have attempted to align definitions here where possible, and focus on relatively recent results that may be more comparable. In addition, while the decennial Census is not a sampling survey, it does have known differential undercounts, specifically of Black and Hispanic people and of children under 4 years old \citep{ohare_differential_2019}. In addition, specific concerns have been raised about the accuracy of the 2020 Census, conducted during the COVID-19 pandemic and amidst political polarization around questions of ethnicity and citizenship \citep{potok_2020_2021}. The availability of response options, for example in race/ethnicity and sex/gender questions, shapes results and may introduce response biases. Any inaccuracies would affect the weights here, particularly if they reflect differential inaccuracies across the demographic variables or between over- and under-represented states.

Using Census rather than American Community Survey (ACS) data allows the analysis to avoid the need to estimate uncertainty from sampling, but it limits the demographic variables to those conducted in the Census short form. Using the survey data would allow more granular trend analysis and potentially analysis of other factors---including income and employment status \citep{united_states_census_bureau_developers_2025} or other measures of rural/urban geography \citep{sanders_rural_2025}---or reduce bias for some demographics \citep{anderson_sampling-based_2000}. The decennial Census also limits analysis to elections at the end of a reapportionment cycle (i.e., comparing the Census demographics to the election in the same year, with apportionment based on the previous Census). This would be at the peak of the so-called ``creeping malapportionment'' of the House and Electoral College \citep{gaines_apportionment_2009} that occurs between instances of redistricting. This may lead to some overstatement of distortions in the House and, to a lesser extent, the Electoral College, compared to other years.

\subsection{Further Research}

This study may provide a basis to examine the complex phenomena of electoral distortion and misrepresentation that arise in the United States, both empirically through models of elections and policy-making and theoretically through notions of representation \citep{alexander_representation_2019}. Indeed, it may also serve as a baseline against which studies of the distortive effects of, for example, the winner-take-all EC or particular election contexts may be judged. These additional features of the system interact with the distortions identified here, potentially exacerbating or reducing disparities; this warrants further study. This may illuminate surprising results if one aspect of the system is changed without changing others, as noted by \citet{thomas_estimating_2013}. In addition, summary measures of distortion could be computed for demographic variables as a whole, as has been done for states \citep{cervas_legal_2020}; differences from those computed for states, which were small, would be interesting.

Further research to improve these metrics and the results derived from them can be conducted using additional data sources. Using representative surveys like ACS could identify the distortions for other demographic groups or for combinations of demographic groups. For example, the underrepresentation of urban non-white Americans may be even more extreme than the product of the weights shown here. Using longitudinal data on a longer or more granular time scale could also reveal more details about the persistence or trends of these patterns, as well as help disentangle the effect of ``creeping malapportionment'' between redistricting compared from perpetual distortions \citep{gaines_apportionment_2009}.

\subsection{Conclusion}

These metrics provide clear numerical and visual ways of analyzing the distortive effects of the key bodies in U.S. national elections. This analysis demonstrates the consistent over- and under-representation of certain groups in the Senate and Electoral College at the expense of others, in particular white and rural populations. Given the history of these bodies, this cannot be viewed as an incidental feature of the system \citep{baker_rural_1955,dupont_distorting_2024}. As state interests per se become perhaps less distinct \citep{alexander_representation_2019}, the distortion of other demographic groups with more identifiable interests \citep[see, e.g.,][]{hoffman_illegitimate_1996} becomes an even more salient feature of the Senate and Electoral College than its weighting of the votes of certain states. Indeed, it is a core feature that displays remarkable persistence and aligns with some historic inequities. Clearly, then, it is likely to shape the formation of parties and platforms, the selection of candidates, and the prospects for various elections and policies. Quantifying, visualizing, and communicating these distortions is a key step to understanding this power and determining, democratically, whether the system needs to be reformed in order to live up to the ideal of true representativeness laid out in \citet{noauthor_gray_1963}: ``The conception of political equality ... can mean only one thing---one person, one vote.''

%%%

\section{Disclosure statement}\label{disclosure-statement}

The author has no conflicts of interest to declare.

\section{Data Availability Statement}\label{data-availability-statement}

Data and relevant code have been made available at the following URLs: https://bit.ly/Elec-Analysis and https://doi.org/10.5281/zenodo.17187083 (live GitHub repository and persistent copy of repository at time of submission); https://bit.ly/Elec-Weights (interactive R Shiny app).

\bibliography{main.bib}
\end{document}